\pdfoutput=1
\documentclass[12pt,preprint]{aastex}

\shorttitle{Sheeley, Rouillard}
\shortauthors{Tracking Streamer Blobs Into the Heliosphere}

\begin{document}

\title{Tracking Streamer Blobs Into the Heliosphere}

\author{N. R. Sheeley, Jr. \&
A. P. Rouillard\footnote{College of Science, George Mason University, Fairfax VA 22030}}
\affil{Space Science Division,
Naval Research Laboratory, Washington DC 20375-5352;
neil.sheeley@nrl.navy.mil,~alexisrouillard@yahoo.co.uk}

\begin{abstract}
In this paper, we use coronal and heliospheric images from the STEREO spacecraft to track streamer
blobs into the heliosphere and to observe them being swept up and compressed by the fast wind from
low-latitude coronal holes.  From an analysis of their elongation/time tracks, we discover a `locus
of enhanced visibility' where neighboring blobs pass each other along the line of sight and their
corotating spiral is seen edge on.  The detailed shape of this locus accounts for a variety of
east-west asymmetries and allows us to recognize the spiral of blobs by its signatures in the
STEREO images:  In the eastern view from STEREO-A, the leading edge of the spiral is visible as a
moving wavefront where foreground ejections overtake background ejections against the sky and then
fade.  In the western view from STEREO-B, the leading edge is only visible close to the Sun-spacecraft
line where the radial path of ejections nearly coincides with the line of sight.  In this case, we
can track large-scale waves continuously back to the lower corona and see that they originate
as face-on blobs.

\end{abstract}


\section{INTRODUCTION}

In 1996--1997, we used the Large-Angle Spectrometric Coronagraph (LASCO) on the {\it Solar and
Heliospheric Observatory} to observe equatorial coronal streamers edge on and to track fragments
of coronal material as they disconnected from the cusps of the streamers and moved radially outward
through the 2--30$R_\odot$ field of view \citep{SHE_97,WANG_98}.  These ejections accelerated through
the corona and eventually reached nearly constant speeds in the range of 300--400 $km~s^{-1}$, as if
they were being carried passively outward in the solar wind.  We called these ejections, `streamer
blobs'.

Now, we use Sun Earth Connection Coronal and Heliospheric Investigation (SECCHI) instruments
(COR2, HI1, HI2) on the {\it Solar Terrestrial Relations Observatory} (STEREO) twin spacecraft,
A and B, to observe streamers face on and to track their detached blobs much farther into the
heliosphere.  COR2 is a Sun-centered coronagraph with a field of 2.0--15.0$R_\odot$.  HI1 and HI2
are offpointing Heliospheric Imagers.  HI1 has a 20${^\circ}$ field centered 13.2${^\circ}$ east
of the Sun for STEREO-A and west of the Sun for STEREO-B.  HI2 has a 70${^\circ}$ field centered
53.4${^\circ}$ east of the Sun for STEREO-A and west of the Sun for STEREO-B.  Together, these
instruments provide low latitude fields of view that range from 2.0$R_\odot$ to $\sim$90${^\circ}$
from the Sun looking eastward from STEREO-A and westward from STEREO-B. \cite{HOW_08} have
provided a more detailed description of these instruments.

In the edge-on views, streamer blobs lie close to the sky plane and move outward at a rate of about
4--6 $day^{-1}$ \citep{WANG_98}.  Their elongation/time tracks are nearly parallel.  Most of the tracks
fade beyond the outer edge of the HI1 field of view, probably due to the decrease in density that
accompanies their expansion and to their increasing distance from the Thomson sphere \citep{VOU_06}.
Some of these tracks may refer to different parts of the same three-dimensional structure seen along
the line of sight \citep{SHE_09}.

By comparison, streamer blobs have a much different appearance when they are observed face on.
In the lower corona, they appear as arch shaped features, as if they were miniature flux ropes
seen from the side \citep{SHE_09}.  Their elongation/time tracks are organized into systematic
patterns that converge in STEREO-A and diverge in STEREO-B, as if the ejections originated from
a fixed longitude on the rotating Sun \citep{SHE_08a, ROU_08,ROU_10a}.  We found that the fixed
longitudes correspond to the north-south segments of coronal streamers at the leading edges of
low-latitude coronal holes.  Recently, \cite{WOOD_10} regarded a large, bow-shaped HI2-A wave
to be the signature of a corotating interaction region (CIR), and empirically traced it back to a
Carrington map of coronal intensity.  Consistent with our own studies, the bow-shaped wave
projected nicely onto a curved segment of the streamer at the leading edge of a low-latitude
coronal hole.

Despite the implication that the STEREO images show streamer blobs being swept up by the fast
wind from low-latitude holes, many questions remain about our interpretation of the observations.
Why are we able to track blobs continuously back to the Sun in STEREO-B, but not in STEREO-A?
Why does the leading STEREO-B track tend to be directed about 73$^\circ$ from the sky plane?
Why do STEREO-A blobs remain visible far beyond the Thomson sphere and then disappear suddenly
after passing other blobs along the line of sight?  Why do we seldom see STEREO-A blobs more
than 60$^\circ$ from the sky plane?

In this paper, we describe a `locus of enhanced visibility', which helps us to answer these
questions.  We introduce the locus in Section 2 and use this locus to interpret STEREO
observations in Section 3.  We summarize these results and discuss their significance in
Section 4, and include their mathematical description in Section 5 (the Appendix).
 
\section{ELONGATION/TIME TRACKS OF EJECTED BLOBS}

In this section, we consider ejections that are released from a fixed point on the
rotating Sun, and we ask what the evolving pattern would look like from the STEREO-A and STEREO-B
spacecraft.  Referring to Figure~1, we relate the elongation angle, $\alpha$, to the radial
distance, r, for a given elevation angle, $\delta$, that the ejection makes with the sky plane.
Applying the Law of Sines to the triangle in Figure~1, we obtain the track equation:
\begin{equation}
r/a = \frac{sin\alpha}{cos(\alpha-\delta)}
\end{equation}
which can be rewritten as
\begin{equation}
tan\alpha = \frac{\rho~cos\delta}{1-\rho~sin\delta}~,   
\end{equation}
where the normalized distance, ${\rho}$, is given by $\rho = r/a$.

As discussed previously \citep{SHE_08a,ROU_08}, when the time dependence of the normalized radial
distance, $\rho$, is considered, equation (2) gives the elongation angle $\alpha$ as a function of
time for each value of the elevation angle, $\delta$.  For example, if the Sun did not rotate and
the ejection moved outward from the Sun toward point P at a constant speed, $v_r$, then we could
substitute the simple relation $\rho = v_{r}t/a$ into equation (2) to obtain the time dependence
of the elongation angle, $\alpha$.  However, the Sun rotates at a synodic angular rate
$\omega$~$\approx$~0.233 rad $day^{-1}$ (corresponding to a period of approximately 27.0 days),
and the initial acceleration is not over until the ejection reaches a radial distance,
${\rho_0}~{\sim}20{R_\odot}/a$.  In this case, ${\rho}$ is given by:
\begin{equation}
\rho = \rho_0 + (\frac{v_r}{a})[t - t_{0}(\delta)],
\end{equation}
provided that $t$ is greater than the starting time $t_{0}(\delta)$.  For simplicity, we consider
ejections that lie close to the ecliptic plane and neglect the 7.25$^\circ$ tilt of the Sun's axis.
In this case, we can set $t_{0}(\delta) = (\pi/2 + \delta)/\omega$, so that the starting time at
$\rho = \rho_0$ will be 0 for an ejection that is directed 90${^\circ}$ behind the sky plane in a
direction opposite to the observer.  We need to go back this far because the STEREO-A elongation/time
tracks start there and show such distant `backside' ejections.

In the westward view from STEREO-B, the elevation angle,
$\delta$, is also defined to be positive in front of the sky plane and negative behind the sky
plane.  Consequently, for STEREO-B, we choose $t_{0}(\delta) = (3\pi/2 - \delta)/\omega$ so that
$t_{0}({\delta})$ continues to increase as the ejection site rotates through the western hemisphere.
With these starting times, equation (3) gives the radial distance, $\rho$, as a linear function
of $\delta$ at each time $t$, and describes a series of rotating spirals like the one that is plotted
in Figure~1.

\subsection{STEREO-A}
Figure~2 shows a STEREO-A plot of the elongation angle, $\alpha$, versus time for values of
$\delta$ ranging from -89${^\circ}$ to +89${^\circ}$, and a radial speed $v_{r} = 330~km~s^{-1}$.
Each track is plotted with a narrow solid line until it grazes the common envelope for a day or so
and then is plotted with a dotted line as it leaves the envelope and bends horizontally.  We use
dotted lines to emphasize the decreased visibility of the corresponding solar ejections after they
leave this envelope and are no longer seen edge-on.  The envelope itself is shown with a thick solid
line which lasts about 18 days, corresponding to the time for the Sun to rotate
180${^\circ}$ ($\pi/\omega = 13.5$ days) plus the Sun-A transit time ($a/v_{r}~{\approx}$ 5 days).

Figure~3 compares an elongation/time map, obtained at a position angle of 100${^\circ}$ during 2008
January 17--31 (upper panel), with the calculated tracks from Figure~2 (middle panel).  The STEREO-A
elongation/time map consists of three sections merged together to span the COR2
($0.5{^\circ}-4.0{^\circ}$), HI1 ($3.2{^\circ}-23.3{^\circ}$), and HI2 ($18.4{^\circ}-88.4{^\circ}$)
fields of view.  After transforming the offpointed HI1 and HI2 images into a Sun-centered polar
coordinate system, radial strips at a particular position angle ($100{^\circ}$ in this case) are
extracted and stacked chronologically to form a rectangular map of intensity difference as a
function of time (horizontal axis) and elongation angle (vertical axis).  These STEREO-A maps are
crossed by a background of downward sloping tracks, which correspond to stars or the remnants of
star images that escaped the differencing used to remove them from the HI2 images.  As we will see
later in Figure~5, the west-to-east motion of the stars produces upward slanting tracks in the
STEREO-B elongation/time maps.  Finally, in the top panel, the vertical bar on January 24 corresponds
to a data gap.  Additional details of the elongation/time maps and their construction are described
elsewhere \citep{SHE_99,SWW_07,SHE_08b,ROU_08,ROU_09,DAV_09}.

In Figure~3, the tracks of interest are the ones of positive slope, which resemble the calculated
tracks and correspond to the motions of streamer blobs moving outward from the Sun.  In transferring
the calculated maps from Figure~2 to the middle panel of Figure~3, we omitted tracks with values of
$\delta$ equal to $70{^\circ},~80{^\circ}$, and $~89{^\circ}$ because tracks with $\delta$ greater
than $60{^\circ}$ do not occur in the top panel of Figure~3.  Such missing high-$\delta$ tracks are a
characteristic of the STEREO-A observations, as we will discuss later.  

The lower panel shows the result of superimposing the calculated tracks (middle panel)
and the observed elongation/time map (upper panel).  Not only do the calculations accurately reproduce
the envelope, but they also reproduce most of the individual tracks.  For a given value of
Sun-spacecraft distance, $a$, the quality of the fit depends on the speed, $v_r$ ($330~km~s^{-1}$),
and the Sun's synodic angular rotation rate, ${\omega}$ ($0.233~rad~day^{-1}$).  But it does not depend
on the value of the assumed normalized radius, $\rho_0$, which disappeared when the calculated pattern
was shifted in time to match the observed pattern.  Allowing for a two-day section that was cropped from
the start of the observed maps, the entire pattern extends for 17 days.

\subsection{STEREO-B}  
Figure~4 shows a STEREO-B plot of elongation angle $\alpha$ versus time for values of $\delta$ ranging
from 89${^\circ}$ to 0${^\circ}$, again calculated for a radial speed $v_{r} = 330~km~s^{-1}$.  The five
tracks with $\delta~{\geq}~75{^\circ}$ graze the envelope twice at points that become increasingly closer
together, and are plotted with dashed lines.  The remaining tracks do not intersect the envelope and are
plotted with narrow solid lines.  Again, we plot the envelope with a thick solid line.  Whereas the
STEREO-A envelope lasted for about 18 days (corresponding to half the solar rotation time plus the Sun-A
transit time), this STEREO-B envelope lasts only about 5 days corresponding to the Sun-B transit time
($a/v_r$).

Figure~5 compares an elongation/time map of COR2-B, HI1-B, and HI2-B images obtained at a position
angle of 268${^\circ}$ during 2008 May 23-29 (upper panel), with the corresponding section of
Figure~4 (middle panel).  As before, their superposition is in the bottom panel.  As we found
previously \citep{SHE_08a,ROU_08}, this western view from STEREO-B is quite different from the
eastern view from STEREO-A with most of the STEREO-B tracks spreading apart.  This behavior is
reproduced in the calculations.  In this particular comparison, the observed tracks span a range
of $\delta$ running from 15${^\circ}$ to 75${^\circ}$.  In fact, the track of largest $\delta$ lies
between the last dashed curve ($\delta$ = 75${^\circ}$) whose intersections with the envelope nearly
coalesce and the first narrow solid curve ($\delta$ = 70${^\circ}$), which just misses the envelope.

\subsection{The locus of enhanced visibility}

We have defined the family of elongation/time tracks that results when ejections are released from
a fixed point on the rotating Sun, and we have seen that these tracks have a common envelope.  In
Appendix 5.1, we deduce the equation of this envelope and show that it corresponds to the locus of
points where the spiral of ejections is seen edge on.  In a
($\rho$,~$\delta$) polar coordinate system, this locus is described by a quadratic equation
\begin{equation}
{\rho}^2 -{\rho}~sin{\delta}~{\pm}~{\nu}~cos{\delta}~=0~,
\end{equation}
where ${\nu}~=~{v_r}/{\omega}a$; the minus sign applies to STEREO-A and the plus sign applies to STEREO-B.  

It is easy to characterize the roots of this equation.
For STEREO-A, there is only one positive, real root given by
\begin{equation}
\rho~=~\frac{sin{\delta}+\sqrt{sin^2{\delta}+4{\nu}~cos{\delta}}}{2}~.
\end{equation}
For STEREO-B, the there are no positive, real roots unless $\delta$ exceeds a critical angle, ${\delta_c}$,
given by
\begin{equation}
 cos{\delta_{c}}~=~\sqrt{(2{\nu})^2+1}-2{\nu}~.
\end{equation}
For a nominal solar wind speed of $330~km~s^{-1}$, $\nu \approx 0.8$, and $\delta_c$ is about
73${^\circ}$.  Thus, in STEREO-B, blobs can pass in front of each other only if they originate
far from the sky plane where $\delta~\ge~\delta_c$.  In this case, there are two positive, real
roots given by
\begin{equation}
{\rho}~=~\frac{sin{\delta}{\pm}\sqrt{sin^2{\delta}-4{\nu}~cos{\delta}}}{2}~.
\end{equation}

We have plotted these relations in Figure~6.  For STEREO-A, the locus lies left of the Sun-spacecraft
line in the region of negative X.  For STEREO-B, the two `roots' join continuously together in the
region of positive X and combine with the STEREO-A locus to form a closed curve with a bean-like shape.
(The link between STEREO-A and STEREO-B would not be continuous if we had used the slightly different
radial distances of these two spacecraft from the Sun.  But for an idealized observer that can look
both east and west, the curve is continuous.)

Because the Thomson scattered intensity increases
where the blobs line up, we have called this bean-shaped curve the `locus of enhanced visibility'.
As we will see in the next section, this locus is useful for tracking streamer blobs through
the STEREO fields of view, especially in combination with a knowledge of the location of the
Thomson sphere \citep{VOU_06}.

In Figure~6, we have represented the Thomson sphere ($\rho$ = sin$\delta$) by a dashed circle.  Also,
we have plotted Sun-centered spirals at intervals of 45${^\circ}$ as thin solid lines to simulate
the corotating spiral of compressed blobs.  In the region of negative X where STEREO-A observes, the
locus of enhanced visibility lies outside the Thomson sphere, as expected from equation (5) where
$\rho$ is always greater than sin$\delta$.  However, in the region of positive X where STEREO-B
observes, the locus lies inside the Thomson sphere as expected from equation (7) where
both roots are less than sin$\delta$.

Equation (7) provides an even more stringent constraint on the locus as seen from STEREO-B.  One
value of $\rho$ is less than $sin{\delta}/2$ and the other value is greater than $sin{\delta}/2$.
This means that these two positive roots correspond to segments of the STEREO-B curve that
respectively lie inside and outside a small circle given by $\rho=sin{\delta}/2$.  This circle
is dotted in Figure~6.  As shown in Appendix 5.2, this circle divides the sky into an inner region
where background ejections appear to outrun foreground ejections against the sky, and an outer
region where foreground ejections appear to outrun background ejections.

Thus, in STEREO-B, a radial ejection either does not intersect the locus of enhanced visibility
at all, or it intersects the locus in two places.  In the latter case, a blob would become aligned
with other ejections twice on its way out from the Sun - first inside the dotted circle where
ejections closer to the sky plane would appear to pass it, and again outside the circle where the
blob would appear to overtake background ejections.  However, this idealized behavior would happen
for only a small range of sky-plane elevations, $\delta$, greater than about 73$^\circ$; for real
blobs with finite angular extents, we can regard the locus as being approximately radial most of
the way outward from the Sun.  Even ejections that miss the locus would still be closely confined
between the locus and the Thomson sphere (provided that they lie on the front side of the Sun),
and therefore frontside ejections would remain visible until they move beyond the Thomson sphere.
By comparison, a streamer blob would encounter the locus of enhanced visibility only once in
STEREO-A; this location would be outside the dotted circle where foreground ejections always pass
background ejections along the line of sight.

With these new results, we are prepared to interpret STEREO observations of streamer blobs
as they are compressed into a corotating spiral by fast wind from low-latitude coronal holes.

\section{OBSERVATIONS}

In the previous section, we showed that the STEREO elongation/time tracks of streamer blobs are
consistent with a model of ejections moving radially outward from a fixed longitude on the rotating
Sun.  Also, we found that the envelopes of these tracks correspond to a bean-shaped locus where
ejections pass in front of each other along the line of sight and their corotating spiral is seen
edge on.  In STEREO-A, the locus does not correspond to the track of any particular ejection, but
is a wavefront where individual blobs surge forward to lead the cloud and then fade.  In STEREO-B,
the locus extends nearly radially outward from the Sun about 73$^\circ$ from the sky plane, causing
streamer blobs that are ejected in this direction to remain visible most of the way outward from
the Sun.  Next, we use observations from STEREO-A and STEREO-B to verify these conclusions.

Figure~7 compares HI2-A images (top rows) with a corresponding elongation/time map obtained at a
position angle of 91${^\circ}$.  In the upper row, the white arrows indicate the outward motion of
a compressed streamer blob at the leading edge of the cloud of ejections. The corresponding track
is indicated by the white arrows in the bottom panel.  The blob fades after 0409 UT on
2008 January 27 and gives up the lead to another blob indicated by black arrows in the second row.
The track of this second blob is indicated by black arrows in the bottom panel.  In this way,
each blob surges forward to briefly take the lead and to make its brief contribution to the envelope
of tracks.  This wave of `billowing blobs' marks the passage of the corotating spiral as it sweeps
through the eastern hemisphere.

Figure~8 provides a similar comparison during 2008 January 28--30, as the last detectable streamer
blob (shown by the white arrows) moves forward to take the lead from a prior blob (shown by black
arrows).  By 0609 UT on January 30, similar ejections at higher and lower latitudes have given the
wave a large bow shape with small-scale ripples along its flanks.  This is the curved wave that
\cite{WOOD_10} tracked back to a coronal streamer winding around the leading edge of a low-latitude
coronal hole.

Figure~9 shows a similar comparison of STEREO-B observations.  Here, the elongation/time tracks appear
continuous from 1 AU all the way back to the lower corona.  They correspond to single ejections rather
than a wave of contributions as shown in the STEREO-A images.  The first ejection (white arrows) was
directed $73^\circ$ out of the sky plane and second ejection (black arrows) was directed $60^\circ$
out of the plane.  Their tracks move toward each other in the HI1 field of view and away again in the
HI2 field, just as we found in Figure~4 for tracks close to the critical angle.  Apparently, these
blobs remained closely aligned along our line of sight until they reached the middle of the HI2-B
field of view.  Like the HI2-A wave in Figure~8, these HI2-B waves in Figure~9 have a rippled
fine structure, apparently reflecting the contributions of individual blobs.
 
Figure~10 shows an ejection that can be followed continuously through the fields of COR2-B,
HI-B, and HI2-B during 2008 June 20-25.  It began as an unremarkable face-on blob, indicated by
the arch-shaped feature in the upper left image.  However, this initially unimpressive
feature evolved into a pronounced azimuthal wave as it moved through the HI1-B field, where we
suppose that it was compressed by the fast wind from the coronal hole behind it.  The ejection
is finally visible as a large, curved wave in the HI2-B field where it disrupted the tail of comet
Boattini before sweeping past Earth.  Because we are able to track this impressive feature
continuously back to the Sun, we know that it did not begin in a dramatic event that somehow
escaped our view.  Rather, it gained its impressive stature gradually in the HI1-B field
as material was swept up and compressed by the fast wind from the coronal hole.

As described previously \citep{SHE_08a,SHE_08b,ROU_08,ROU_09}, we can fit the well-defined
STEREO-B track with equations (2) and (3), corresponding to a radial trajectory at a constant
speed from a location about 20$R_\odot$ from the Sun.  In this case, we find that
${\delta}~{\approx}~64{^\circ}$ and $v_{r}~{\approx}~370~km~s^{-1}$.  Thus, the elevation angle
of this ejection was only $10{^\circ}$ less than the critical angle of $74{^\circ}$ needed for
this $370~km~s^{-1}$ ejection to graze the STEREO-B envelope.  This closeness to the envelope
allowed us to detect the blob continuously from the Sun to 1 AU and to determine its origin
as a face-on blob.

Figure~11 compares this STEREO-B elongation/time map with the corresponding map obtained
for STEREO-A.  Because comet Boattini was located just south of the Sun-Earth line, approximately
equidistant from STEREO-A and STEREO-B, its STEREO-A elongation angle should also be about
${\delta}=64{^\circ}$.  But the corresponding track in STEREO-A is not visible in COR2-A or
HI1-A, and it was hardly visible in HI2-A until the comet arrived on June 24.  This is typical
of streamer blobs ejected farther than about $60^\circ$ from the STEREO-A sky plane.  Referring
to Figure~6, we suppose that the reduced visibility of such streamer blobs is caused by their
unfavorable locations far from the sky plane, well inside the Thomson sphere, and far from
the locus of edge-on views.  Figure~11 also illustrates that comets, as well as streamer
ejections, can serve as tracers of CIRs.

\section{SUMMARY AND DISCUSSION}

Using a model in which streamer blobs are ejected from a fixed point on the rotating Sun, we
derived an equation for their elongation/time tracks as seen from STEREO-A and STEREO-B.
We showed that these two sets of tracks had envelopes similar to structures seen in the
elongation/time maps of observed data.  We derived equations for those envelopes and found
that they described the places that streamer blobs pass each other along the line of sight.
These are also the places that one observes the corotating spiral of compressed blobs edge on.
We solved these equations and obtained a bean-shaped curve in the Sun's equatorial plane.
Because the Thomson scattered intensity is increased where blobs pass each other along the
line of sight, we called this curve `the locus of increased visibility'.  Like the Thomson
sphere \citep{VOU_06,THER_08}, this locus provides a way of interpreting the visibility of
streamer blobs as they pass through the STEREO fields of view.

Although this locus describes the places that the corotating spiral is seen edge on (i.e its
leading edge), the locus itself is not a spiral.  Rather it is a bean-shaped curve whose
east-west asymmetry accounts for the differing visibility of blobs in STEREO-A and STEREO-B
images.  Prior to the launch of STEREO, it was generally supposed that solar ejections would
fade out soon after passing the Thomson sphere \citep{VOU_06}.  In STEREO-A, the locus of
enhanced visibility lies outside the Thomson sphere, creating a crescent-shaped region between
these two boundaries where the ejections remain visible.  However, for sky-plane elevations,
$\delta$, greater than about $60^\circ$, streamer blobs are seldom visible because they are
located far from the sky plane and well inside both the Thomson sphere and the locus of
enhanced visibility for most of their trip outward from the Sun.

In the STEREO-B field of view the locus is oriented almost radially, and ejections close to this
direction ($\delta{\approx}73^\circ$) remain continuously visible most of the way outward from
the Sun.  Even for lower values of $\delta$, the radial paths of frontside ejections lie close to
the locus and to the Thomson sphere.  Consequently, they remain visible until they pass the
Thomson sphere.  This explains why we are able to track streamer blobs continuously
outward from the Sun in the STEREO-B images, but not in STEREO-A images.

In Figures~7--11, we used STEREO observations to illustrate these ideas.
Figures~7 and 8 showed streamer blobs replacing each other at the leading edge of the cloud and
contributing to the envelope of their elongation/time tracks.  Now, we know that this ongoing
replacement is a consequence of the motion of blobs past the locus of enhanced visibility where
the corotating spiral is seen edge on.  Figure~9 showed STEREO-B elongation/time tracks
approaching each other close to the Sun and then separating again in the HI2-B field of view,
as we now expect for streamer blobs that are ejected close to the nearly radial segment of the
locus of enhanced visibility.

Figure~10 showed a large HI2-B wave that we tracked continuously back to a face-on streamer blob
in the COR2-B field of view, and Figure~11 showed that we could not track the corresponding HI2-A
wave back to its origin.  It is easy to understand why we could not track the feature in STEREO-A;
it was ejected $\sim$64$^\circ$ from the STEREO-A sky plane and spent most of its time poorly
visible inside the Thomson sphere.  However, it is not easy to understand how such a large HI2
wave could be generated by compressing a single streamer blob.  Perhaps this observation is
reminding us that streamer blobs are just tracers of the more massive sweeping that occurs when
fast wind runs into the streamer.  This continuous flow would not be visible in our running
difference images, but its compressed plasma might eventually be seen in the HI2 field
of view, as \cite{WOOD_10} supposed.  In this sense, one could think of the streamer as being
composed of many unresolved blobs.  We also need to appreciate that the unambiguous
association applies only at one position angle along the HI2 wave, and that neighboring parts
of the wave front may be associated with other blobs.

The fast wind from the coronal hole is essential for tracking these ejections into the heliosphere.
Discrete ejections occur everywhere along the streamer belt at the rate of about $4-6~day^{-1}$,
\citep{SHE_97,WANG_98}, but if those ejections are not compressed by fast wind from a coronal hole,
then their radial expansion will cause them to fade toward the end of the HI1 field of view.
Consequently, the surviving blobs originate just ahead of the coronal hole where the streamer is
distorted into a north-south segment and appears face on
\citep{SHE_08a, ROU_08, ROU_09, ROU_10a,WOOD_10}.

The same argument applies to the continuous flow from a streamer, whose face-on segment is
compressed into a corotating interaction region (CIR) \citep{GOS_HUND_77,BUR_95}.  \cite{TAP_09}
recently focused on the CIR, calculating its expected structure and looking for its observational
signatures in the STEREO and the Solar Mass Ejection Imager (SMEI) fields of view.  The centerpiece
of their study was the `spine' or `backbone' that lies along the upper edge of the pattern of
STEREO-A elongation/time tracks.  They identified this feature with the leading edge of the CIR
where the density spiral is seen edge on, and they regarded the more rapidly rising tracks to be
`irregularities or waves' ahead of the CIR.  Likewise, they regarded the STEREO-B tracks to be
signatures of `knots and waves' that had not yet formed a stable leading edge.

It is instructive to consider why the blobs become more visible when they cross the bean-shaped
locus and pass each other along the line of sight.  As we mentioned in Section 2.3, the coalignment
of two blobs would increase the Thomson scattered intensity simply by adding more scattering material
along the line of sight.  However, if the blobs themselves were compressed and squashed along the
spiral, then their edge-on orientation would provide an even greater contribution than what would be
obtained for randomly oriented blobs or uncompressed blobs passing each other along the line of sight.
Images like those shown in Figures~7 and 10 give us the impression that the blobs are being compressed
as they pass through the HI1 field of view, and it is tempting to suppose that the dominant effect
comes from the alignment of their flat edges as the spiral is seen edge on.

In STEREO-A, the edge-on views occur for only a day or two because the radially moving blobs cross
the bean-shaped locus nearly normal to its boundary.  However, in STEREO-B, the coalignment lasts
for almost the entire 5-day transit to 1 AU, until the blobs finally peel away from each other and
reveal large waves in the HI2-B field of view.

In this paper, we limited our attention to streamer blobs being swept up by high-speed streams from
low-latitude coronal holes, and we did not consider the streamer disconnection events
\citep{WANG_99,SIM_97} and `streamer blowout' coronal mass ejections \citep{HOW_85,SWW_07,SHE_07},
which we also expect to be swept up by the high-speed streams.  As in previous studies, we were led
to the idea that streamer blobs are tracers, but now with face-on blobs revealing the formation of
CIRs in much the same way that edge-on blobs reveal the acceleration of the solar wind \citep{SHE_97}.
It is only a short step further to suppose that the main portion of the solar wind originates in open
magnetic field regions (whose coronal divergence determines the wind speed), and that everything else
(blobs, streamer detachments, `streamer blowout' CMEs, and the unresolved streamer itself) is just
tracer material ejected into that wind.

\section{APPENDIX}

\subsection{The Envelope Equation}
In this section, we derive an equation for the envelope of the family of elongation/time
tracks, and we show that it corresponds to the locus of places that the corotating spiral
of streamer blobs is seen edge on.  We begin by writing equation (2) in the form
\begin{equation}
tan{\alpha}~=~\frac{{\rho}(t,{\delta})~cos{\delta}}{1-{\rho}(t,{\delta})~sin{\delta}}~,
\end{equation}
where ${\rho}(t,{\delta})$ is given by equation (3).  It is well known that the envelope of such a
family of curves is obtained by combining this equation with its partial derivative with respect to
$\delta$.  Taking this derivative and keeping $\alpha$ and $t$ constant, we find that
\begin{equation}
{\rho}^2 -{\rho}~sin{\delta} + {\rho^\prime}~cos{\delta}~=0,
\end{equation}
where ${\rho^\prime}=-({v_r}/a)~{t^\prime_0}({\delta})$.  In this case,
${\rho^\prime}={\pm}{v_r}/{\omega}a$, with the minus sign for STEREO-A and the plus sign for
STEREO-B.  For convenience, we define ${\nu}={v_r}/{\omega}a$, and write
${\rho^\prime}={\pm}{\nu}$.  Together these relations define the envelopes for STEREO-A and
STEREO-B.  In Section 5.3 of this appendix, we will provide an alternate definition in terms
of the elongation angle, $\alpha$ and the normalized radial distance, $\rho$.

The same relations are obtained for the locus of points where the spiral of ejections is seen edge on.
To see this, we return to Figure~1 and look along the line that extends from point A through point P
and beyond.  Our objective is to keep the elongation angle, $\alpha$, constant and ask how the
normalized radial distance, $\rho$ changes as the elevation angle, $\delta$, is varied.  For this
purpose, we simply take the derivative of equation (8) with respect to $\delta$ while keeping
$\alpha$ constant.  The result is again given by equation (9).  So if $\rho$ changes with
$\delta$ as given by this equation, then neighboring ejections will lie along our line of sight from
point A to point P and beyond.  For these ejections to also lie on the spiral, the rate of
change, ${\rho^\prime}$, must be obtained from equation (3), in which case
${\rho^\prime}={\pm}{\nu}$ with the minus sign for STEREO-A and the plus sign for STEREO-B.
Thus, the locus of edge-on views and the envelope of tracks are determined by the same equations
and therefore are the same.

\subsection{Separating the angular speeds of foreground and background ejections}
The dotted circle in Figure~6 divides the heliosphere into an inner region where background ejections outrun
foreground ejections against the sky, and an outer region where foreground ejections outrun
background ejections.  To see this, we first hold $\delta$ constant and calculate the angular rotation
rate $\dot{\alpha}$.  Then, we hold $\alpha$ constant and determine how $\dot{\alpha}$ changes with
$\delta$.  We expect to find that ${\partial}{\dot{\alpha}}/{\partial}{\delta}$ is positive
outside the dotted circle and negative inside it.  We begin with equation (1).  Taking the time
derivative and recognizing that $\dot{\rho}$ = $v_{r}/a$, we find that
\begin{equation}
 \dot{\alpha}~=~{\frac{v_r}{a}}~ \frac{cos^{2}(\alpha-\delta)}{cos{\delta}}
\end{equation}
Thus, for a given direction, $\delta$, a blob reaches its largest angular speed, $\dot{\alpha}$, at
the Thomson sphere where ${\alpha}={\delta}$ and the intensity is greatest. 
  
Next, we hold $\alpha$ constant, and take the logarithmic derivative of equation (10) with respect to
$\delta$.  Combining the result with equation (2), we obtain
\begin{equation}
\frac{1}{\dot{\alpha}} \frac{{\partial}{\dot{\alpha}}}{{\partial}{\delta}}~=~\frac{2\rho-sin\delta}{cos\delta}
\end{equation}
and therefore
\begin{equation}
\frac{{\partial}{\dot{\alpha}}}{{\partial}{\delta}}~=~{\frac{v_r}{a}}~\frac{cos^{2}(\alpha-\delta)}{cos^{2}\delta}~(2\rho-sin\delta).
\end{equation}
From this equation, it is clear that the angular rotation speed $\dot{\alpha}$ increases with $\delta$
outside the circle (where $\rho=sin{\delta}/2$), shown dotted in Figure~6, and decreases with $\delta$
inside that circle.
This means that background ejections appear to overtake foreground ejections inside the dotted circle and
foreground ejections appear to overtake background ejections outside the circle.

\subsection{Elongation angles along the locus of enhanced visibility}

In Section 5.1, we wrote the envelope equation as a relation between normalized radial distance,
$\rho$, and the ejection's elevation angle, $\delta$, out of the sky plane.  Now, we wish to
rewrite this equation as a relation between the elongation angle, $\alpha$, and the normalized
radial distance, $\rho$.  We could obtain the desired relation by simply eliminating $\delta$ from
equations (2) and (4).  However, it is instructive, and perhaps easier,
to take a physical approach.  In this case, we recognize that for a spiral
\begin{equation}
tan({\alpha}-{\delta})~=~\frac{{\partial}{\rho}}{{\rho}~{\partial}{\delta}}~=~\frac{v_r}{{\omega}{r}}~=~\frac{\nu}{\rho}~.
\end{equation}
Solving this equation for $cos({\alpha}-{\delta})$ and then equating it to the value of 
$cos({\alpha}-{\delta})$ given by equation (1), we obtain the desired result:
\begin{equation}
sin{\alpha}~=~\frac{{\rho}^2}{\sqrt{{\nu}^2+{\rho}^2}}~.
\end{equation}

This equation is useful for identifying a point along the envelope in an elongation/time map if
the value of $\rho$ is known or easily calculated.  For example, if we wish to identify
the point in Figure~4 where the two STEREO-B tangents coalesce, we set ${\delta}={\delta_c}$ in
equation (7) and obtain ${\rho}=sin{\delta_c}/2$.  Then, using ${\nu}~{\approx}~0.8$
in equations (6) and (14), we find that ${\alpha}~{\approx}~14.2{^\circ}$, consistent with a location
midway between the tangent points of the dashed curves in Figure~4.

The STEREO/SECCHI data are produced by a consortium of NRL (US), LMSAL (US), NASA/GSFC (US), RAL (UK),
UBHAM (UK), MPS (Germany), CSL (Belgium), IOTA (France), and IAS (France).  In the US, funding was
provided by NASA, in the UK by PPARC, in Germany by DLR, in Belgium by the Science Policy Office, and
in France by CNES and CNRS.  NRL received support from the USAF Space Test Program and ONR.  We are
grateful to our many colleagues in these organizations who made these observations possible.  In
particular, we would like to acknowledge Nathan Rich (NRL) and Tom Cooper (Cornell University) for a
wide variety of programming assistance and Yi-Ming Wang (NRL) for his continued scientific help and
advice.  We have also benefited from helpful scientific discussions with Arnaud Thernisien (USRA) and
Brian Wood (NRL).

\bibliography{ms}
\bibliographystyle{apj}

\clearpage

\begin{figure*}[t]
 \centerline{%
 \includegraphics[clip,scale=0.95, angle=90]{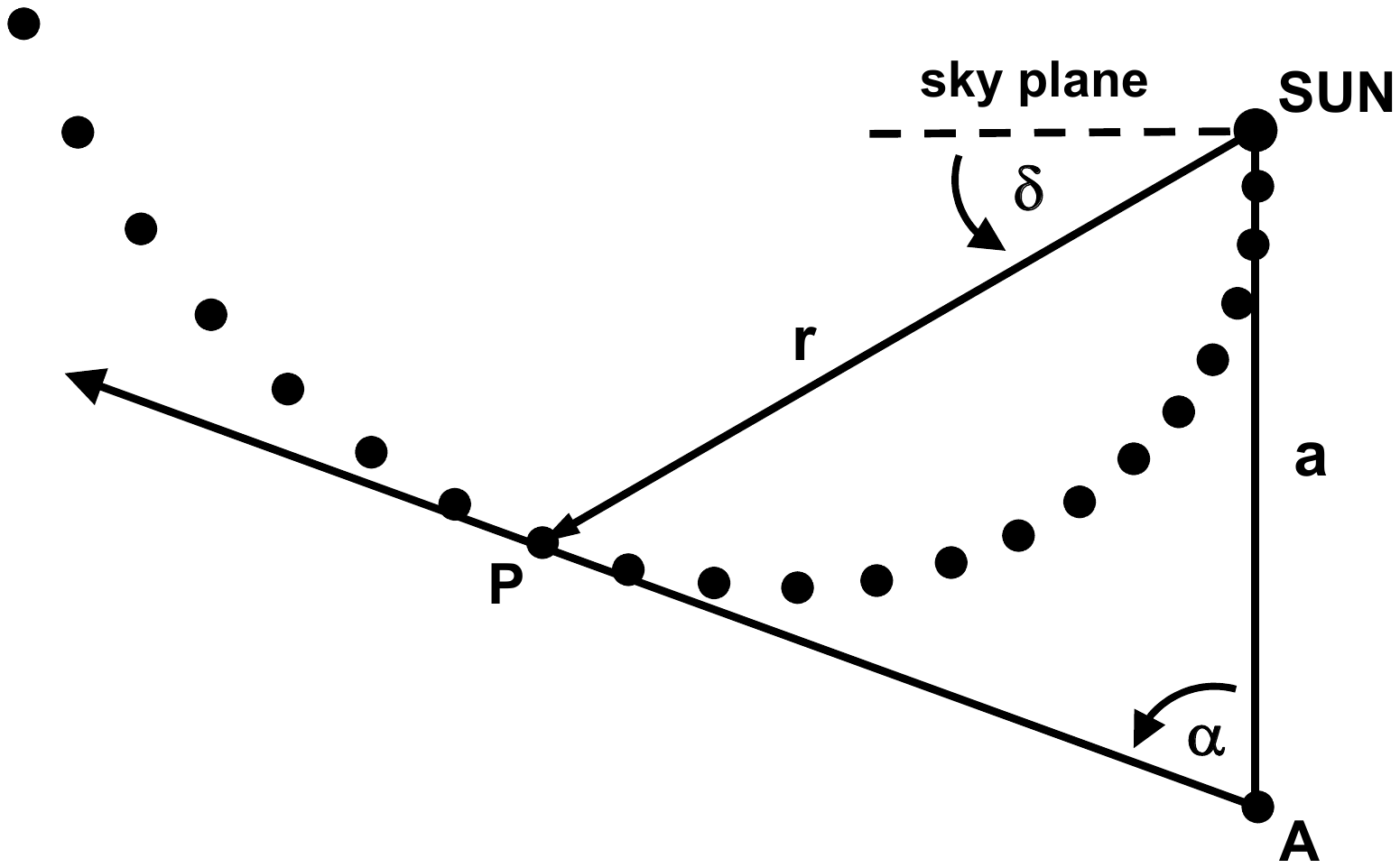}}
 \caption{Sketch, illustrating the geometry of a family of streamer blobs moving radially outward
from a fixed point on the rotating Sun.  Looking $\alpha$ radians east of the Sun, the
observer at point A sees the resulting spiral of ejections edge-on at the point P.}
\end{figure*}

\begin{figure*}[t]
 \centerline{%
 \includegraphics[clip,scale=0.95]{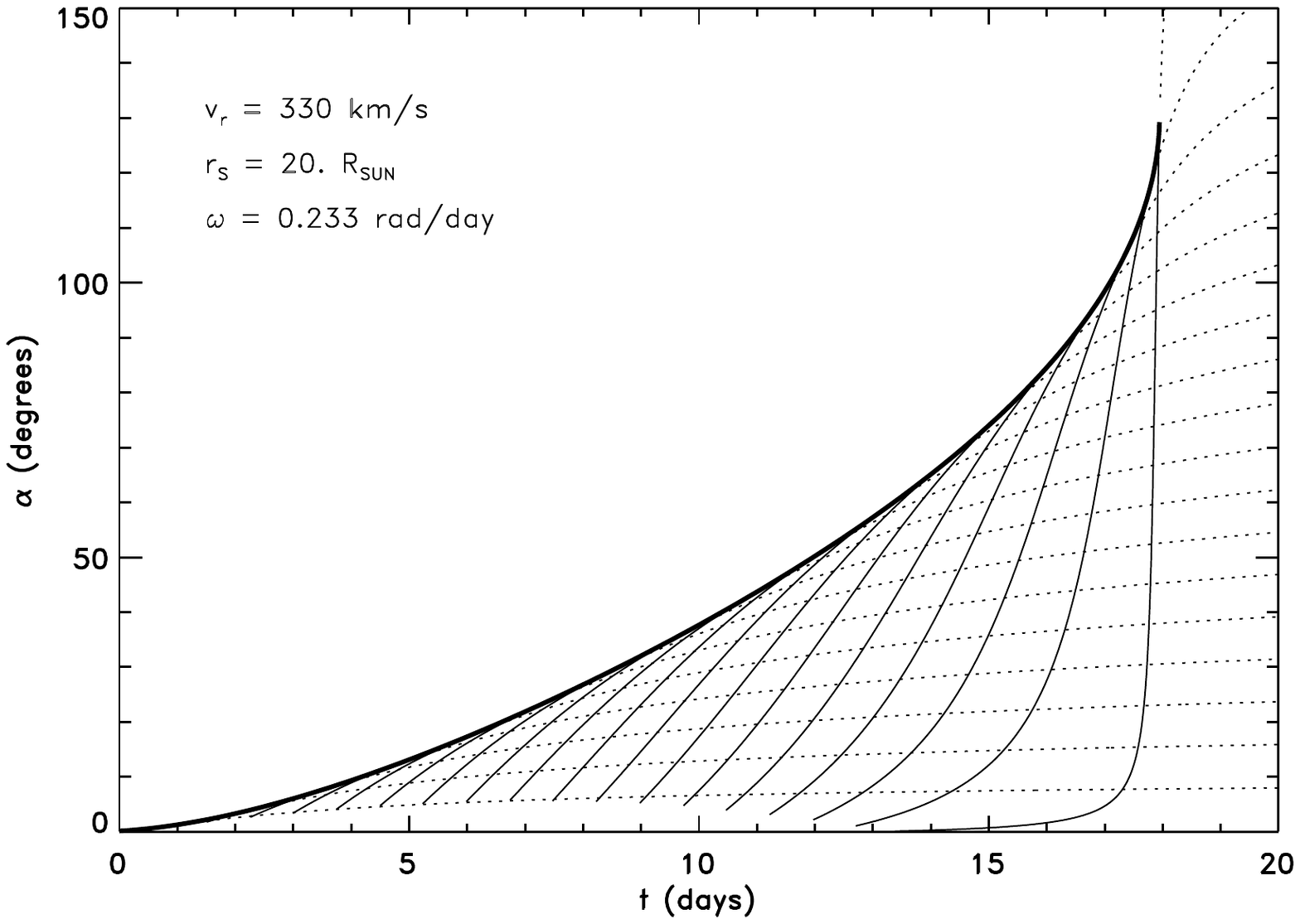}}
 \caption{The eastern view.  Tracks of elongation angle, $\alpha$, versus time, calculated for
sky-plane elevation angles, $\delta$ = $-89{^\circ}$, -80${^\circ}$, -70${^\circ}$,
-60${^\circ}$, ..., 70${^\circ}$, 80${^\circ}$, 89${^\circ}$.  Each ejection originates from a
fixed longitude on a 27-day rotating sphere of radius $r_s$~=~20$R_\odot$ beyond which the radial
speed, $v_r$, has a constant value of 330 km s$^{-1}$.  Each track is plotted with a thin solid line until
it reaches the common envelope, and then with a dotted line to emphasize the decreased visibility of
the corresponding solar ejection.}
\end{figure*}
  
\begin{figure*}[t]
 \centerline{%
 \includegraphics[clip,scale=0.75]{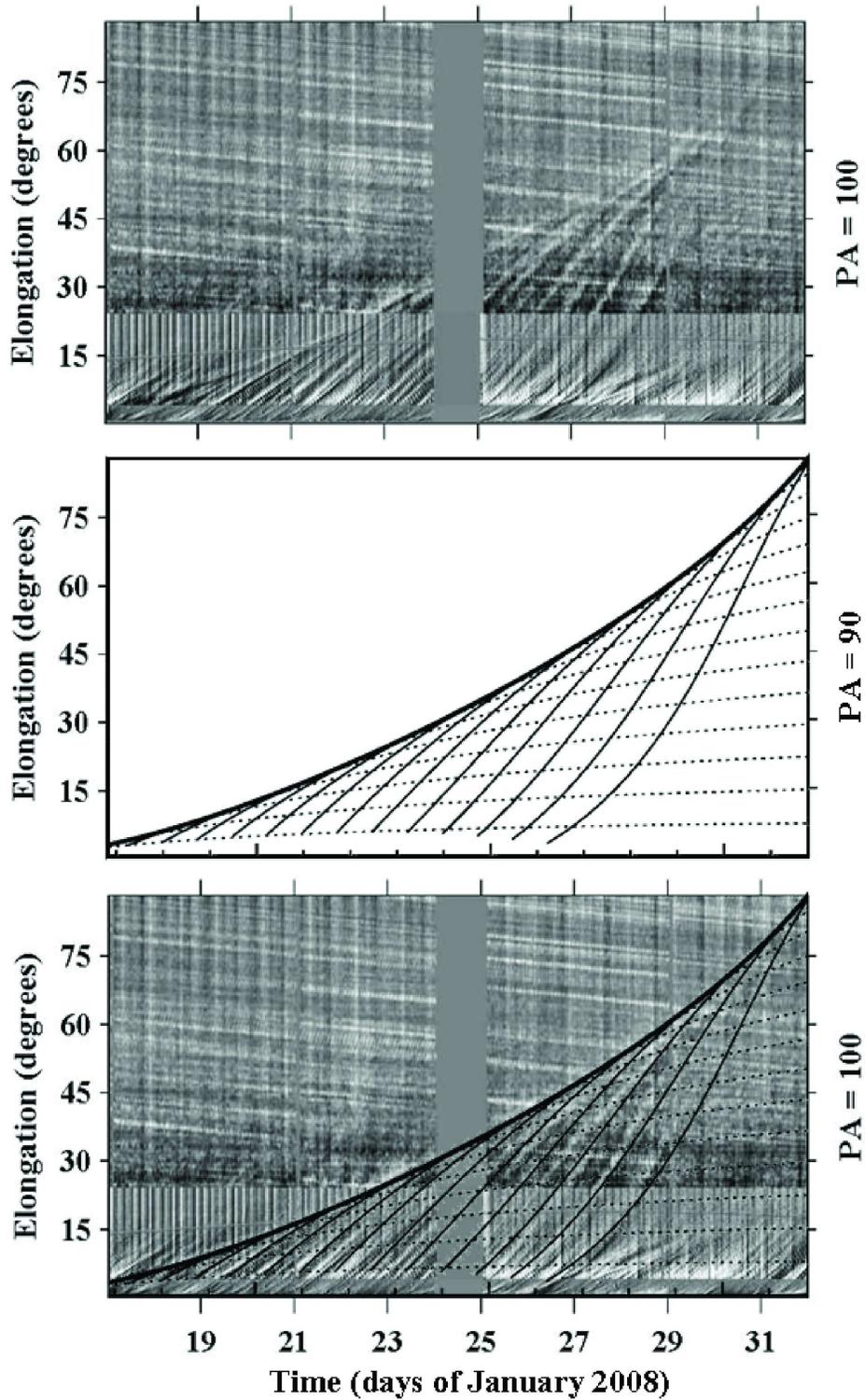}}
 \caption{A comparison between the elongation/time tracks seen from STEREO-A during 2008 January 18--31
(top panel) and the calculated tracks from Fig. 2 with $\delta$ $\leq$ 60${^\circ}$ (middle panel).
As seen in the superimposed images (bottom panel), the observed tracks and their envelope are well
fit by the calculated curves.}
\end{figure*}

\begin{figure*}[t]
 \centerline{%
 \includegraphics[clip,scale=0.95]{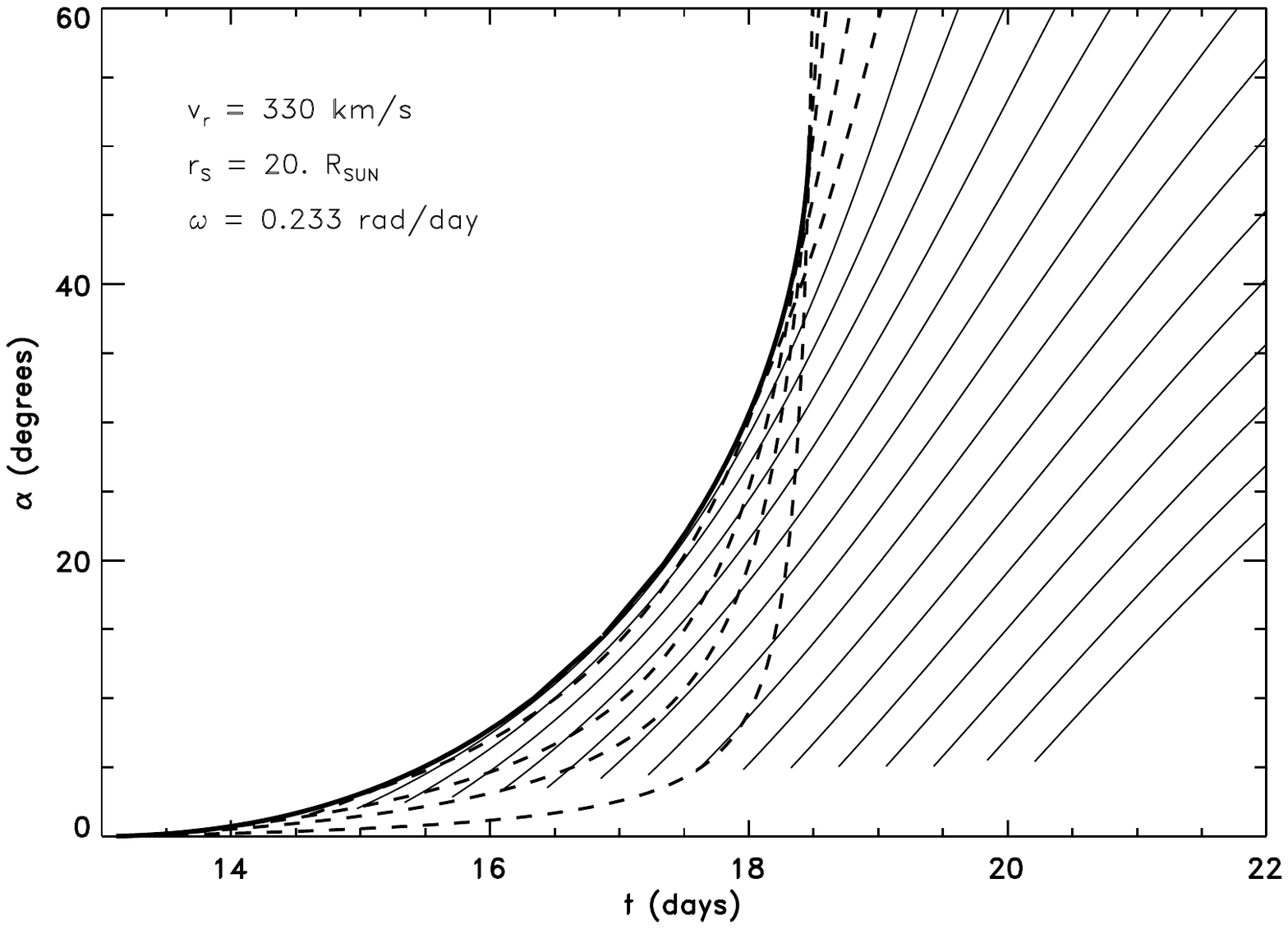}}
 \caption{The western view.  Similar to Fig. 2 except that tracks are plotted for sky-plane
elevation angles, ${\delta}$ = $89{^\circ}$, $87{^\circ}$, $85{^\circ}$, $80{^\circ}$, $75{^\circ}$, $70{^\circ}$, ..., $10{^\circ}$, $5{^\circ}$, $0{^\circ}$.  The five tracks with $\delta$ greater than the critical value,
$\delta_c$~$\sim$~73${^\circ}$, graze the envelope twice and are plotted with dashed lines;
tracks with smaller values of $\delta$ miss the envelope and are plotted with solid lines.}
\end{figure*}

\begin{figure*}[t]
 \centerline{%
 \includegraphics[clip,scale=0.70]{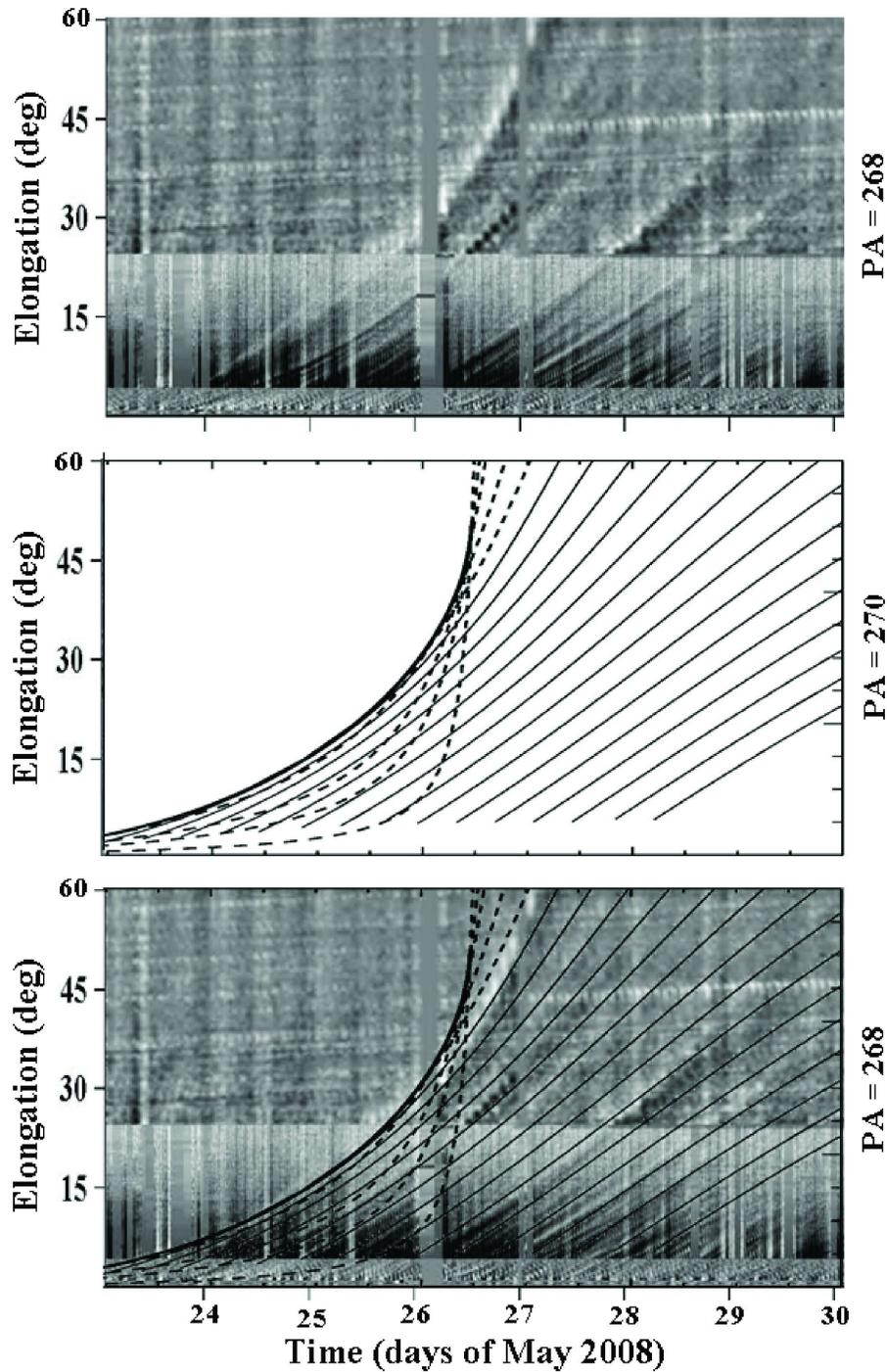}}
 \caption{A comparison between the elongation/time tracks seen from STEREO-B during 2008 May 23--29
(top panel) and the calculated tracks from Fig. 4 (middle panel).  As shown in the superimposed images
(bottom panel), the observed tracks correspond to sky-plane elevations, $\delta$, in the range
(75${^\circ}$, 15${^\circ}$).}
\end{figure*}

\begin{figure*}[t]
 \centerline{%
 \includegraphics[clip,scale=1.4]{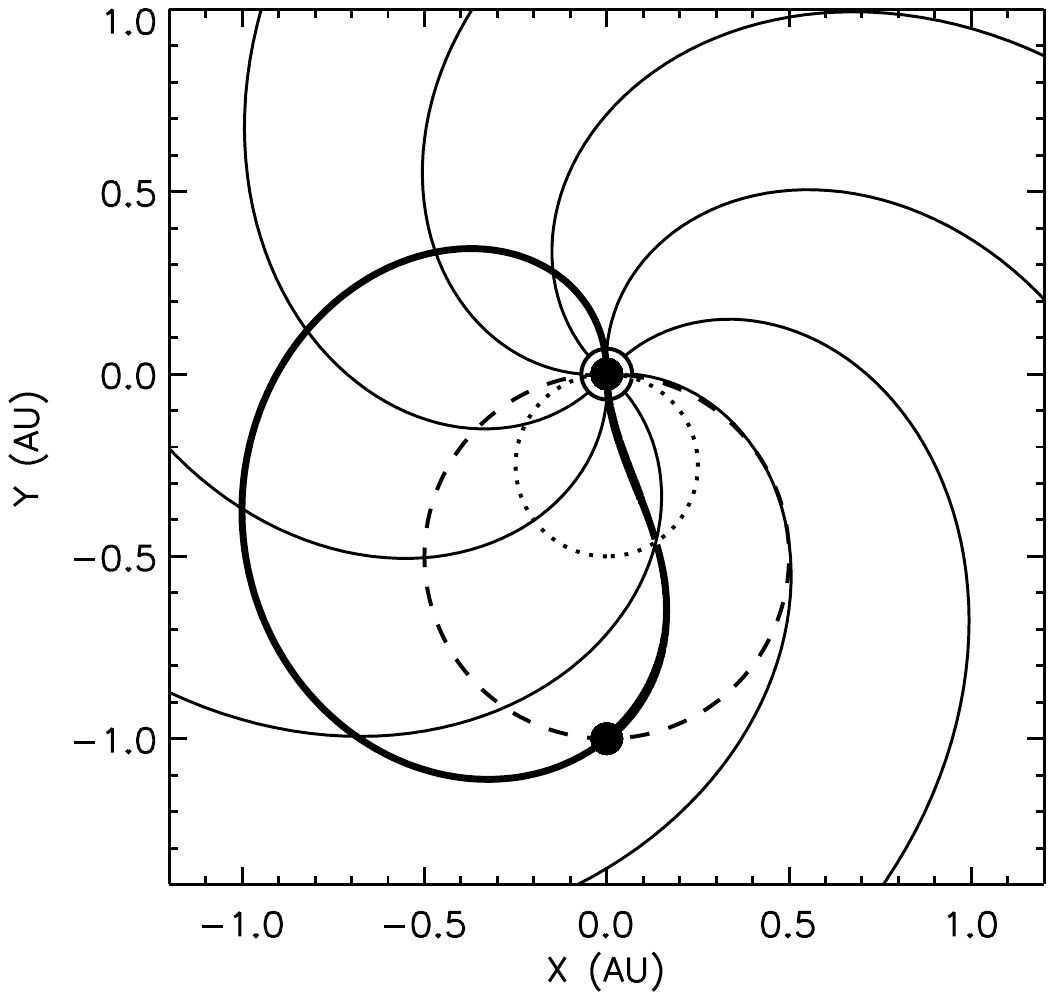}}
 \caption{A polar coordinate view of the Sun's equatorial plane, showing the locus of enhanced visibility
(thick solid curve) and the Thomson sphere (dashed circle).  The Sun is at the origin, surrounded by a
20$R_\odot$ region of acceleration, and the observer is at the point (0, -1).  Spirals of radially moving
material are drawn at 45${^\circ}$ intervals.  The observer sees the leading edge of the spiral moving
along the bean-shaped locus.  The dotted circle separates an inner region, where background ejections pass
foreground ejections against the sky, from an outer region, where foreground ejections pass background
ejections.}
\end{figure*}

\begin{figure*}[t]
 \centerline{%
 \includegraphics[clip,scale=0.75]{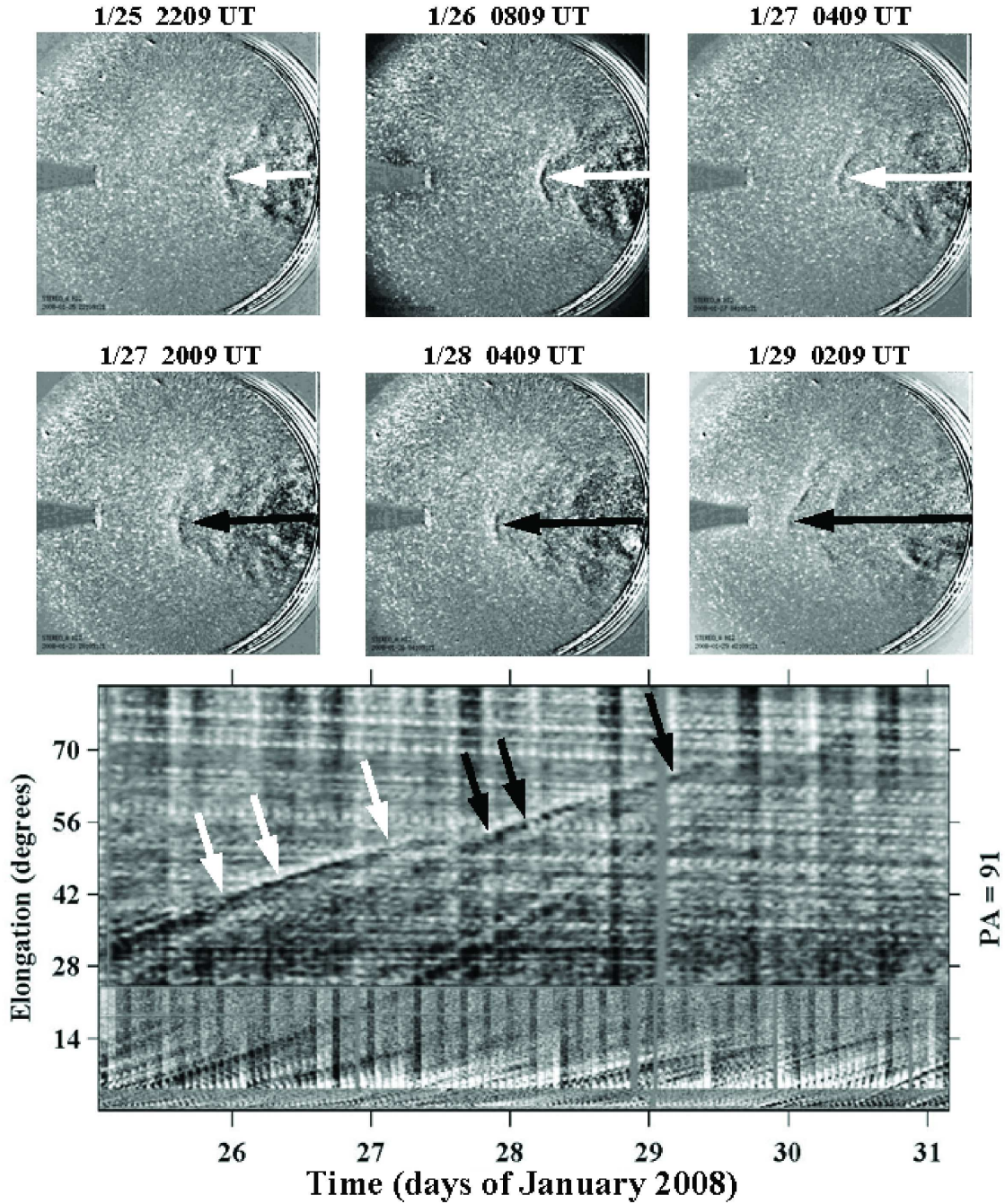}}
 \caption{HI2-A images (top two rows) and a composite COR2-HI1-HI2 elongation/time map (bottom),
showing the replacement of streamer blobs at the leading edge of the cloud.  A compressed blob
(white arrows) surges forward to take the lead before being overtaken along the line of sight by
another blob (black arrows) located farther from the plane of the sky.  The envelope of tracks marks
the leading edge of the corotating spiral as seen from STEREO-A.}  
\end{figure*}

\begin{figure*}[t]
 \centerline{%
 \includegraphics[clip,scale=0.75]{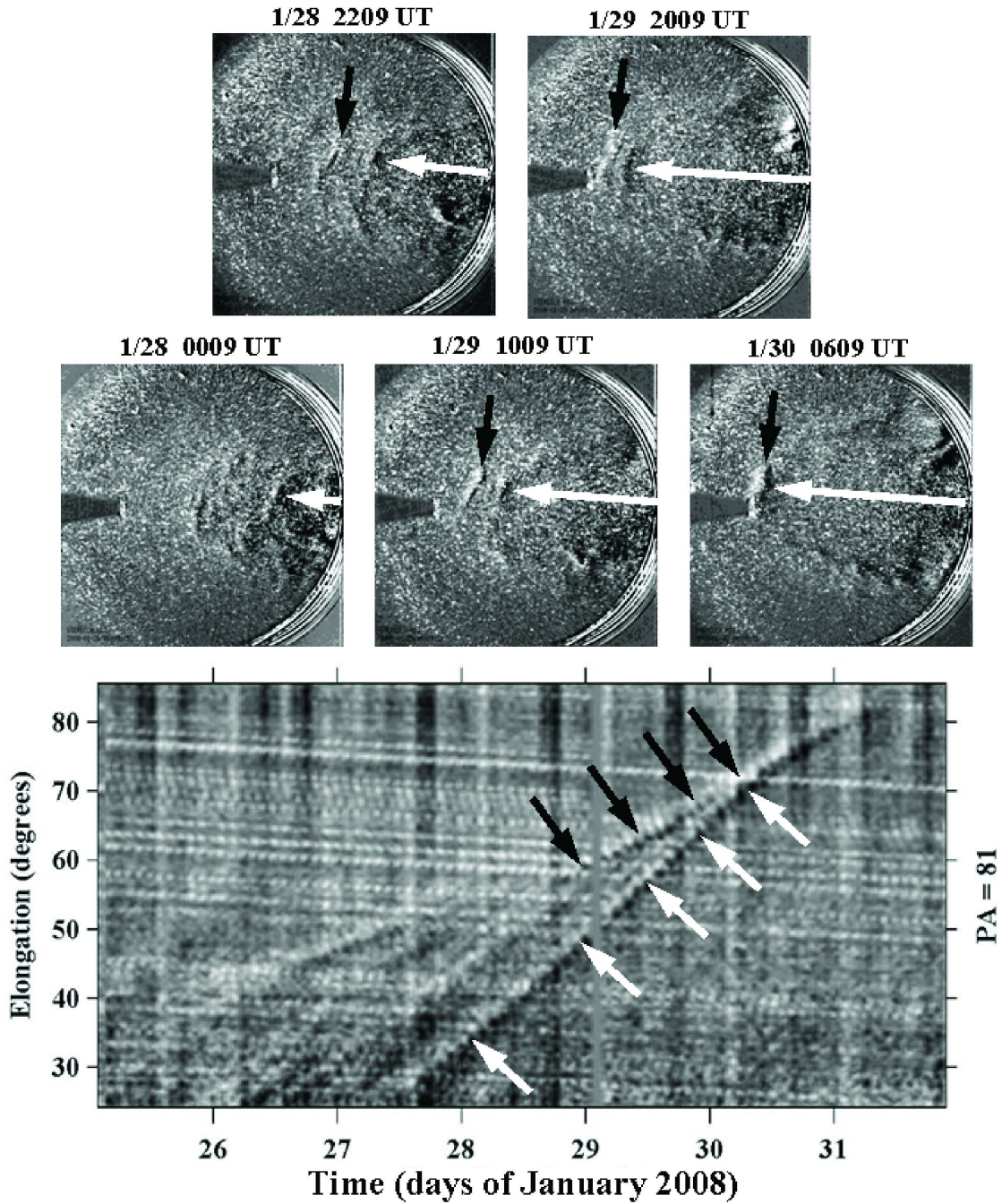}}
 \caption{HI2-A images (alternating between the middle and top panels) and a HI2-A elongation/time
map (bottom panel), showing the last visible ejection (white arrows) moving forward to take the lead
from its most recent leader (black arrows).  Similar replacements occur at adjacent position angles,
producing the bow-shaped wave with the rippled fine structure seen in the last image at 0609 UT on
January 30.}
\end{figure*}

\begin{figure*}[t]
 \centerline{%
 \includegraphics[clip,scale=0.70]{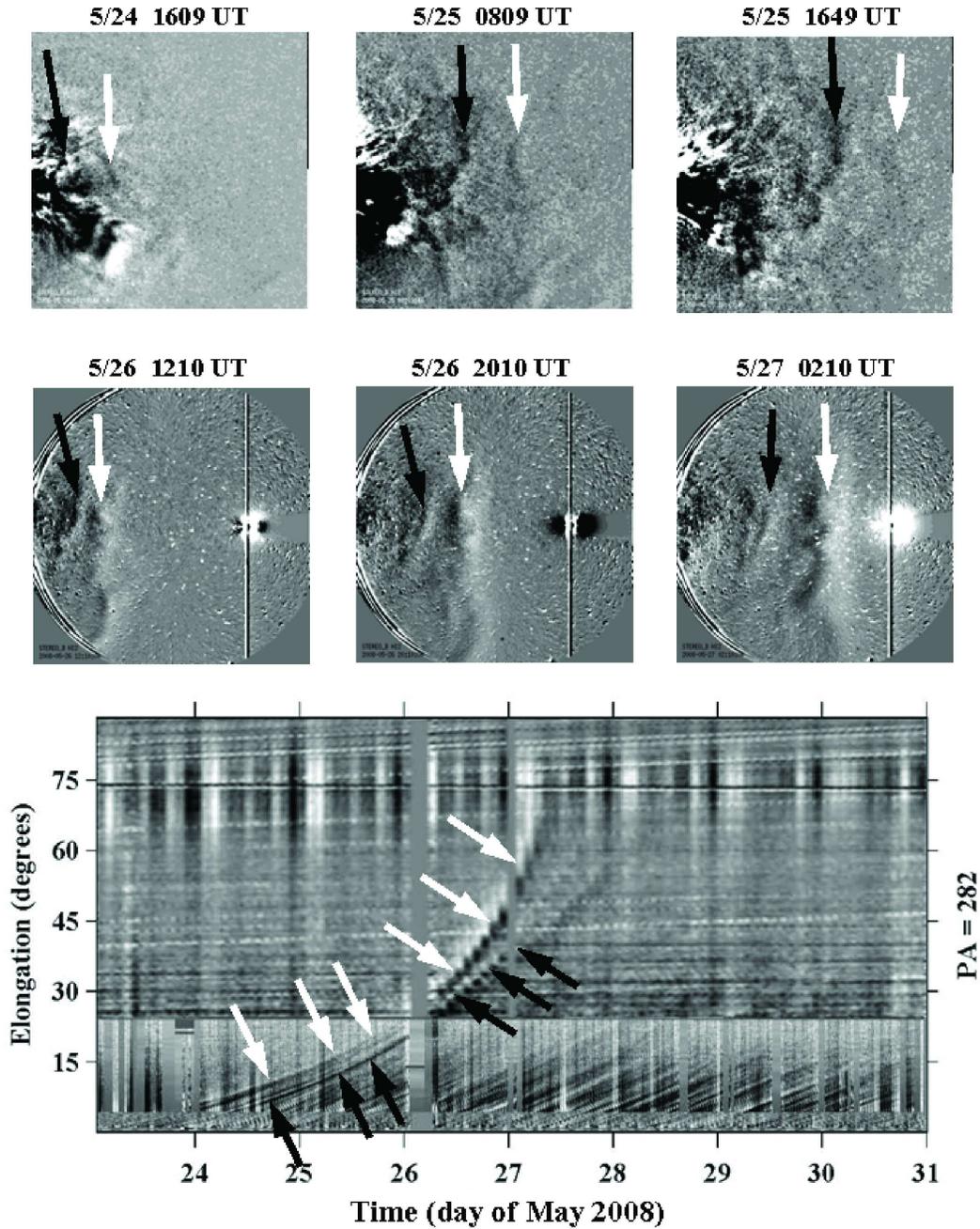}}
 \caption{HI1-B images (top row), HI2-B images (middle row), and a composite COR2-HI1-HI2
elongation/time map (bottom panel), showing streamer blobs that were ejected close to the locus
of enhanced visibility during 2008 May 23--30.  The track with ${\delta}=73^\circ$ (white arrows)
approaches the track with ${\delta}=60^\circ$ (black arrows) in the HI1 field of view and
separates again in the HI2 field.}
\end{figure*}

\begin{figure*}[t]
 \centerline{%
 \includegraphics[clip,scale=0.70]{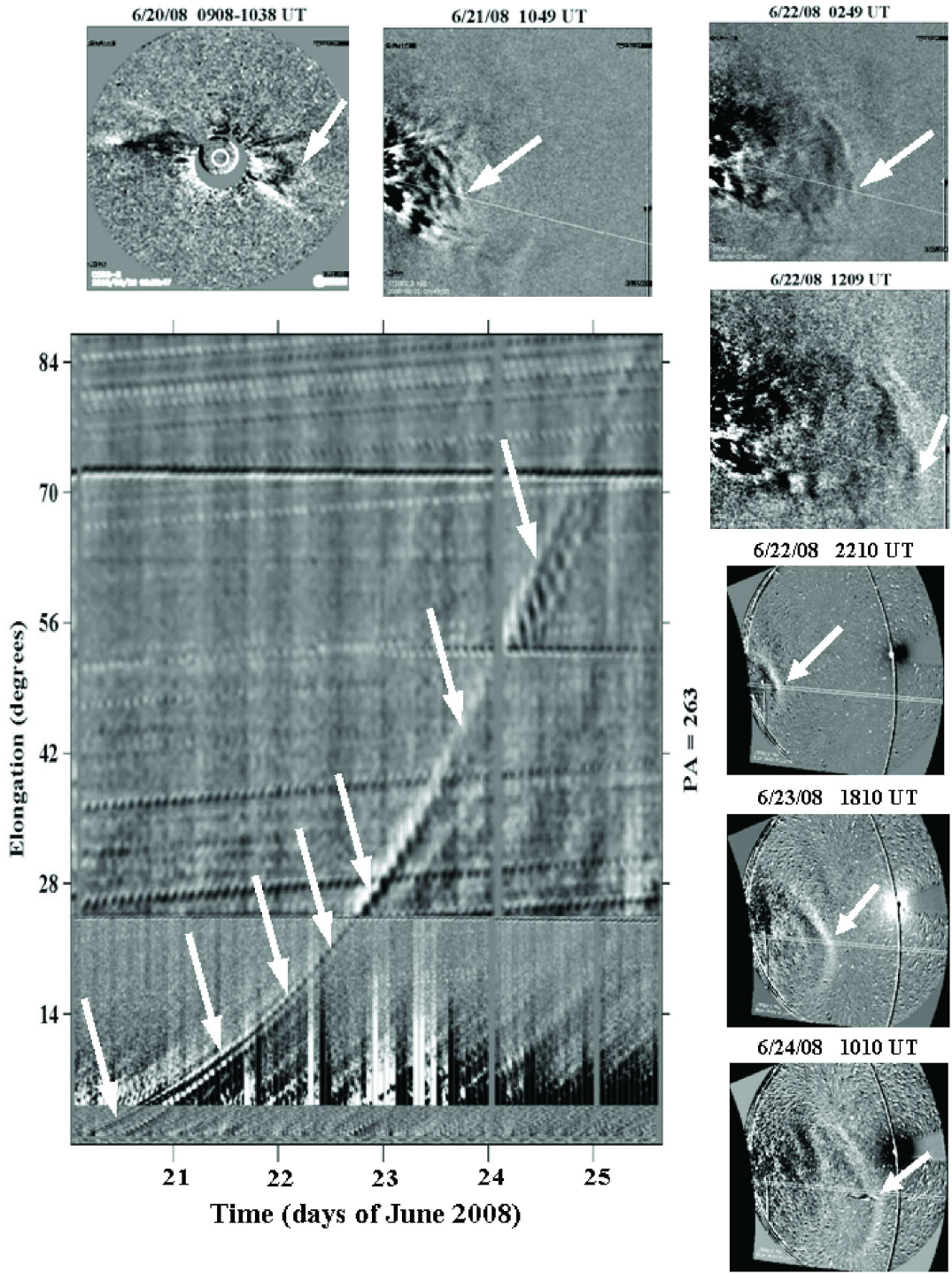}}
 \caption{(clockwise from the upper left) A time sequence of images during 2008 June 20--24, showing
the evolution of a face-on streamer blob (white arrow in the COR2-B image) as it becomes compressed
into an azimuthal structure in the HI1-B field and sweeps past comet Boattini in the HI2-B field.
(lower left) The corresponding elongation/time map, showing the complete track of motion with the
instantaneous positions indicated by white arrows.  In the last image, the radial lines have
been shifted slightly northward from their true location so that they do not obscure the comet tail.}
\end{figure*}

\begin{figure*}[t]
 \centerline{%
 \includegraphics[clip,scale=0.85]{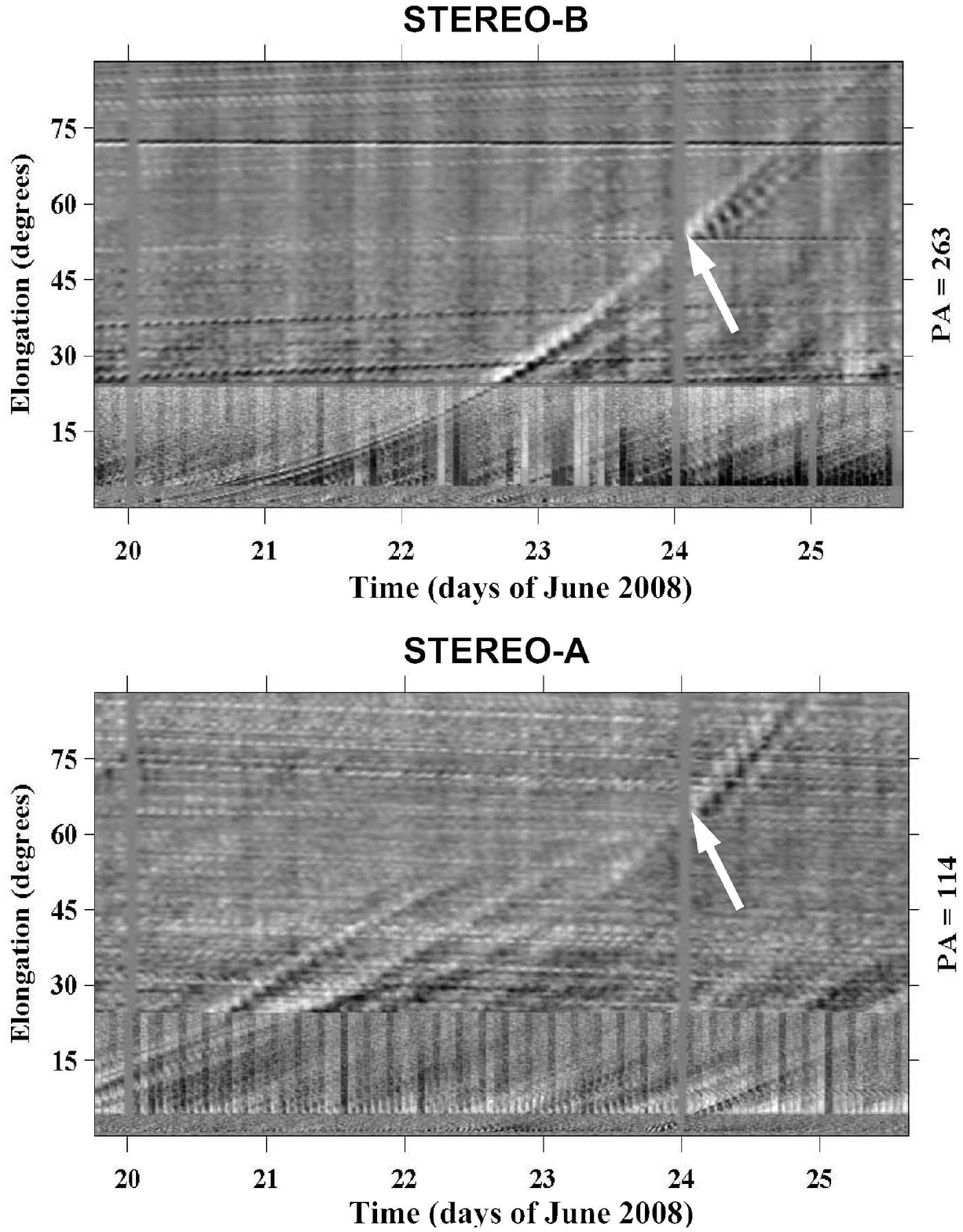}}
 \caption{Comparison of STEREO-A and STEREO-B elongation/time maps during 2008 June 19--25 when the
fast wind from a coronal hole swept up a streamer blob and disturbed the tail of comet Boattini
(white arrows), which was located nearly midway between the two spacecraft on the Sun-Earth line.
Despite the symmetry between the comet's location in these two fields of view, only in STEREO-B can
the wave be traced back to its origin in the lower corona.}
\end{figure*}

\end{document}